\documentclass{article}

\usepackage{arxiv}

\usepackage[utf8]{inputenc} % allow utf-8 input
\usepackage[T1]{fontenc}    % use 8-bit T1 fonts
\usepackage{hyperref}       % hyperlinks
\usepackage{url}            % simple URL typesetting
\usepackage{booktabs}       % professional-quality tables
\usepackage{amsfonts}       % blackboard math symbols
\usepackage{nicefrac}       % compact symbols for 1/2, etc.
\usepackage{microtype}      % microtypography
\usepackage{lipsum}
\usepackage{graphicx}
\usepackage{multicol}

\title{‘Warriors of the Word’ - Deciphering Lyrical Topics in Music and Their Connection to Audio Feature Dimensions Based on a Corpus of Over 100,000 Metal Songs}

\author{
  Isabella~Czedik-Eysenberg \\
  Department of Musicology\\
  University of Vienna, Austria \\
  \texttt{isabella.czedik-eysenberg@univie.ac.at} \\
   \And
  Oliver~Wieczorek \\
  Department of Sociology, esp. Sociological Theory\\
  University of Bamberg, Germany \\
  \texttt{oliver.wieczorek@uni-bamberg.de} \\
  \And
  Christoph~Reuter \\
  Department of Musicology\\
  University of Vienna, Austria \\
  \texttt{christoph.reuter@univie.ac.at} \\
}

\begin{document}
\maketitle

\begin{abstract}
We look into the connection between the musical and lyrical content of metal music by combining automated extraction of high-level audio features and quantitative text analysis on a corpus of 124.288 song lyrics from this genre. Based on this text corpus, a topic model was first constructed using Latent Dirichlet Allocation (LDA). For a subsample of 503 songs, scores for predicting perceived musical \emph{hardness/heaviness} and \emph{darkness/gloominess} were extracted using audio feature models. By combining both audio feature and text analysis, we (1) offer a comprehensive overview of the lyrical topics present within the metal genre and (2) are able to establish whether or not levels of \emph{hardness} and other music dimensions are associated with the occurrence of particularly harsh (and other) textual topics. Twenty typical topics were identified and projected into a topic space using multidimensional scaling (MDS). After Bonferroni correction, positive correlations were found between musical \emph{
hardness} and \emph{darkness} and textual topics dealing with ‘\emph{brutal death}’, ‘\emph{dystopia}’, ‘\emph{archaisms and occultism}’, ‘\emph{religion and satanism}’, ‘\emph{battle}’ and ‘\emph{(psychological) madness}’, while there is a negative associations with topics like ‘\emph{personal life}’ and ‘\emph{love and romance}’.
\end{abstract}

% keywords can be removed
%\and High-level Music Dimensions \and Hardness \and Heaviness \and Darkness
\begin{multicols}{2}
\section{Introduction}
As audio and text features provide complementary layers of information on songs, a combination of both data types has been shown to improve the automatic classification of high-level attributes in music such as genre, mood and emotion \cite{neumayer2007, laurier2008, hu2010, kim2010}. Multi-modal approaches interlinking these features offer insights into possible relations between lyrical and musical information (see \cite{nichols2009, mcvicar2011, yu2019}).

In the case of metal music, sound dimensions like loudness, distortion and particularly \emph{hardness} (or \emph{heaviness}) play an essential role in defining the sound of this genre \cite{berger2005, reyes2008, mynett2013, herbst2017}. Specific subgenres – especially doom metal, gothic metal and black metal – are further associated with a sound that is often described as \emph{dark} or \emph{gloomy} \cite{phillips2009, yavuz2017}.

These characteristics are typically not limited to the acoustic and musical level. In a research strand that has so far been generally treated separately from the audio dimensions, lyrics from the metal genre have come under relatively close scrutiny (cf. \cite{farley2016}). Topics typically ascribed to metal lyrics include \emph{sadness}, \emph{death}, \emph{freedom}, \emph{nature}, \emph{occultism} or \emph{unpleasant/disgusting objects} and are overall characterized as \emph{harsh}, \emph{gloomy}, \emph{dystopian}, or \emph{satanic} \cite{purcell2015, farley2016, taylor2016, podoshen2014, cheung2019}.

Until now, investigations on metal lyrics were limited to individual cases or relatively small corpora – with a maximum of 1,152 songs in \cite{cheung2019}. Besides this, the relation between the musical and the textual domain has not yet been explored. Therefore, we examine a large corpus of metal song lyrics, addressing the following questions:

\begin{enumerate}
\item Which \emph{topics} are present within the corpus of metal lyrics?
\item Is there a connection between characteristic musical dimensions like \emph{hardness} and \emph{darkness} and certain topics occurring within the textual domain?
\end{enumerate}

\section{Methodology}
\label{sec:methodology}

\begin{figure}
  \centering
  \fbox{\includegraphics[width=0.8\linewidth]{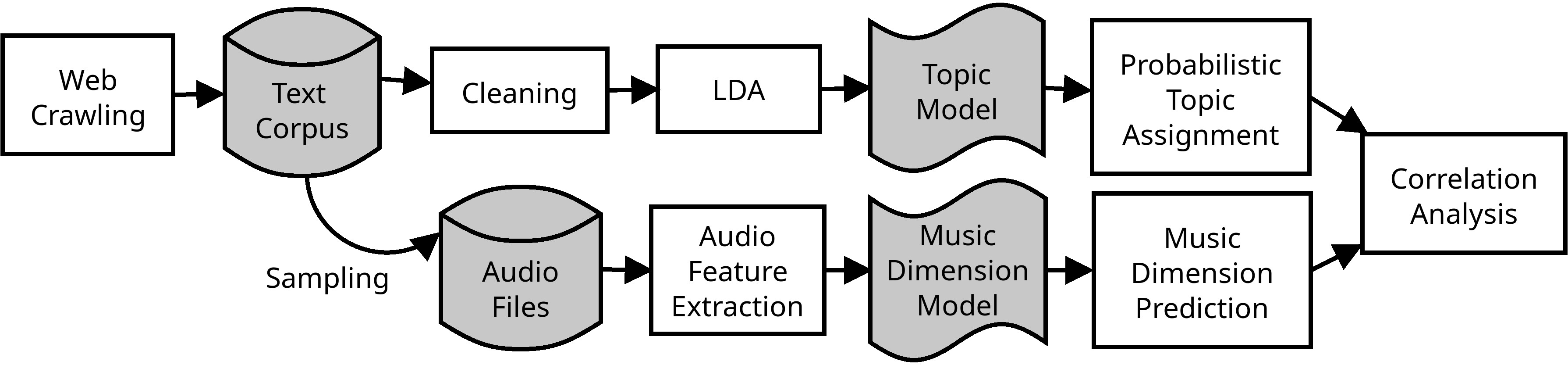}}
  \caption{Processing steps of the approach illustrating the parallel analysis of text and audio features}
  \label{fig:processing_steps}
\end{figure}

In our sequential research design, the distribution of textual topics within the corpus was analyzed using latent Dirichlet allocation (LDA). This resulted in a topic model, which was used for a probabilistic assignment of topics to each of the song documents. Additionally, for a subset of these songs, audio features were extracted using models for high-level music dimensions. The use of automatic models for the extraction of both text as well as musical features allows for scalability as it enables a large corpus to be studied without depending on the process of manual annotation for each of the songs. The resulting feature vectors were then subjected to a correlation analysis. Figure \ref{fig:processing_steps} outlines the sequence of the steps taken in processing the data. The individual steps are explained in the following subsections.

\subsection{Text Corpus Creation and Cleaning}

For gathering the data corpus, a web crawler was programmed using the Python packages \texttt{Requests} and \texttt{BeautifulSoup}. In total, 152,916 metal music lyrics were extracted from \texttt{www.darklyrics.com}.

Using Python’s \texttt{langdetect} package, all non-English texts were excluded. With the help of regular expressions, the texts were scanned for tokens indicating meta-information, which is not part of the actual lyrics. To this end, a list of stopwords referring to musical instruments or the production process (e.g. ‘recorded’, ‘mixed’, ‘arrangement by’, ‘band photos’) was defined in addition to common stopwords. After these cleaning procedures, 124,288 texts remained in the subsample. For text normalization, stemming and lemmatization were applied as further preprocessing steps.

\subsection{Topic Modelling via Latent Dirichlet Allocation}
We performed a LDA \cite{blei2003} on the remaining subsample to construct a probabilistic topic model. The LDA models were created by using the Python library \texttt{Gensim} \cite{rehurek2010}. The lyrics were first converted to a bag-of-words format, and standard weighting of terms provided by the \texttt{Gensim} package was applied.

Log perplexity \cite[p. 4]{coleman2015} and log UMass coherence \cite[p. 2]{roeder2015} were calculated as goodness-of-fit measures evaluating topic models ranging from 10 to 100 topics. Considering these performance measures as well as qualitative interpretability of the resulting topic models, we chose a topic model including 20 topics – an approach comparable with \cite{viola2019}. We then examined the most salient and most typical words for each topic.

Moreover, we used the \texttt{ldavis} package to analyze the structure of the resulting topic space \cite{sievert2014}. In order to do so, the Jensen-Shannon divergence between topics was calculated in a first step. In a second step, we applied multidimensional scaling (MDS) to project the inter-topic distances onto a two-dimensional plane. MDS is based on the idea of calculating dissimilarities between pairs of items of an input matrix while minimizing the strain function \cite{borg2003}. In this case, the closer the topics are located to one another on the two-dimensional plane, the more they share salient terms and the more likely a combination of these topics appear in a song.

\subsection{High-Level Audio Feature Extraction}
The high-level audio feature models used had been constructed in previous examinations \cite{czedik2017, czedik2018}. In those music perception studies, ratings were obtained for 212 music stimuli in an online listening experiment by 40 raters. 

\end{multicols}

\begin{multicols}{2}

Based on this ground truth, prediction models for the automatic extraction of high-level music dimensions – including the concepts of perceived \emph{hardness/heaviness} and \emph{darkness/gloominess} in music – had been trained using machine learning methods. In a second step, the model obtained for \emph{hardness} had been evaluated using further listening experiments on a new unseen set of audio stimuli \cite{czedik2018}. The model has been refined against this backdrop, resulting in an $R^2$ value of 0.80 for \emph{hardness/heaviness} and 0.60 for \emph{darkness/gloominess} using five-fold cross-validation. 

The resulting models embedded features implemented in \texttt{LibROSA} \cite{mcfee2015}, \texttt{Essentia} \cite{bogdanov2013} as well as the \texttt{timbral models} developed as part of the \texttt{AudioCommons} project \cite{pearce2017}.

\subsection{Investigating the Connection between Audio and Text Features}

Finally, we drew a random sample of 503 songs and used Spearman's $\rho$ to identify correlations between the topics retrieved and the audio dimensions obtained by the high-level audio feature models. We opted for Spearman’s $\rho$ since it does not assume normal distribution of the data, is less prone to outliers and zero-inflation than Pearson’s $r$. Bonferroni correction was applied in order to account for multiple-testing.

\section{Results}

\subsection{Textual Topics}

Table \ref{tab:topic_table} displays the twenty resulting topics found within the text corpus using LDA. The topics are numbered in descending order according to their prevalence (weight) in the text corpus. For each topic, a qualitative interpretation is given along with the 10 most salient terms\footnote{Note that the terms are presented in their stemmed form (e.g. ‘fli’ instead of ‘fly’ or ‘flying’).}.

The salient terms of the first topic – and in parts also the second – appear relatively generic, as terms like e.g. ‘know’, ‘never’, and ‘time’ occur in many contexts. However, the majority of the remaining topics reveal distinct lyrical themes described as being characteristic for the metal genre. ‘\emph{Religion \& satanism}’ (topic \#5) and descriptions of ‘\emph{brutal death}’ (topic \#7) can be considered as being typical for black metal and death metal respectively, whereas ‘\emph{battle}’ (topic \#6), ‘\emph{landscape \& journey}’ (topic \#11), ‘\emph{struggle for freedom}’ (topic \#12), and ‘\emph{dystopia}’ (topic \#15), are associated with power metal and other metal subgenres.

\end{multicols}
\begingroup
\renewcommand{\arraystretch}{1.5}
\begin{table}
\caption{Overview of the resulting topics found within the corpus of metal lyrics (n = 124,288) and their correlation to the dimensions \emph{hardness} and \emph{darkness} obtained from the audio signal (see section \ref{sec:results_corr})}
\resizebox{\textwidth}{!}{
\begin{tabular}{ |c|c|c|c|c| } 
 \hline
 \textbf{Topic} & \textbf{Interpretation} & \textbf{Salient Terms (Top 10)} & \textbf{Darkness $\rho$} & \textbf{Hardness $\rho$} \\
 \hline
 1 & \emph{personal life} & know, never, time, see, way, take, life, feel, make, say & -0.195** & -0.262** \\
  \hline
 2 & \emph{sorrow \& weltschmerz} & life, soul, pain, fear, mind, eye, lie, insid, lost, end & 0.042 & -0.002 \\
  \hline
 3 & \emph{night} & dark, light, night, sky, sun, shadow, star, black, moon, cold & -0.082 & -0.098* \\
  \hline
 4 & \emph{love \& romance} & night, eye, love, like, heart, feel, hand, run, see, come & -0.196** & -0.273** \\
  \hline
 5 & \emph{religion \& satanism} & god, hell, burn, evil, soul, lord, blood, death, satan, demon & 0.11* & 0.164** \\
  \hline
 6 & \emph{battle} & fight, metal, fire, stand, power, battl, steel, sword, burn, march & 0.158** & 0.159** \\
  \hline
 7 & \emph{brutal death} & blood, death, dead, flesh, bodi, bone, skin, cut, rot, rip & 0.176** & 0.267** \\
  \hline
 8 & \emph{vulgarity} & fuck, yeah, gon, like, shit, littl, head, girl, babi, hey & 0.075 & 0.056 \\
  \hline
 9 & \emph{archaisms \& occultism} & shall, upon, thi, flesh, thee, behold, forth, death, serpent, thou & 0.115* & 0.175** \\
  \hline
 10 & \emph{epic tale} & world, time, day, new, end, life, year, live, last, earth & 0.013 & 0.018 \\
  \hline
 11 & \emph{landscape \& journey} & land, wind, fli, water, came, sky, river, high, ride, mountain & -0.067 & -0.125* \\
  \hline
 12 & \emph{struggle for freedom} & control, power, freedom, law, nation, rule, system, work, peopl, slave & 0.095* & 0.076 \\
  \hline
 13 & \emph{metaphysics} & form, space, exist, beyond, within, knowledg, shape, mind, circl, sorc & 0.066 & 0.064 \\
  \hline
 14 & \emph{domestic violence} & kill, mother, children, pay, child, live, anoth, father, name, innoc & 0.060 & 0.070 \\
  \hline
 15 & \emph{dystopia} & human, race, disease, breed, destruct, machin, mass, seed, destroy, earth & 0.191** & 0.240** \\
  \hline
 16 & \emph{mourning rituals} & ash, word, dust, stone, speak, weep, smoke, breath, tongu, funer & 0.031 & 0.015 \\
  \hline
 17 & \emph{(psychological) madness} & mind, twist, brain, mad, self, half, mental, terror, urg, obsess & 0.153* & 0.107* \\
  \hline
 18 & \emph{royal feast} & king, rain, drink, fall, crown, sun, rise, bear, wine, color & -0.046 & -0.031 \\
  \hline
 19 & \emph{Rock'n'Roll lifestyle} & rock, roll, train, addict, explod, wreck, shock, chip, leagu, raw & 0.032 & 0.038 \\
  \hline
 20 & \emph{disgusting things} & anim, weed, ill, fed, maggot, origin, worm, incest, object, thief & 0.075 & 0.064 \\ 
 \hline
\end{tabular}
\label{tab:topic_table}
}
\end{table}
\endgroup

\begin{figure}
  \centering
  \fbox{\includegraphics[width=460pt]{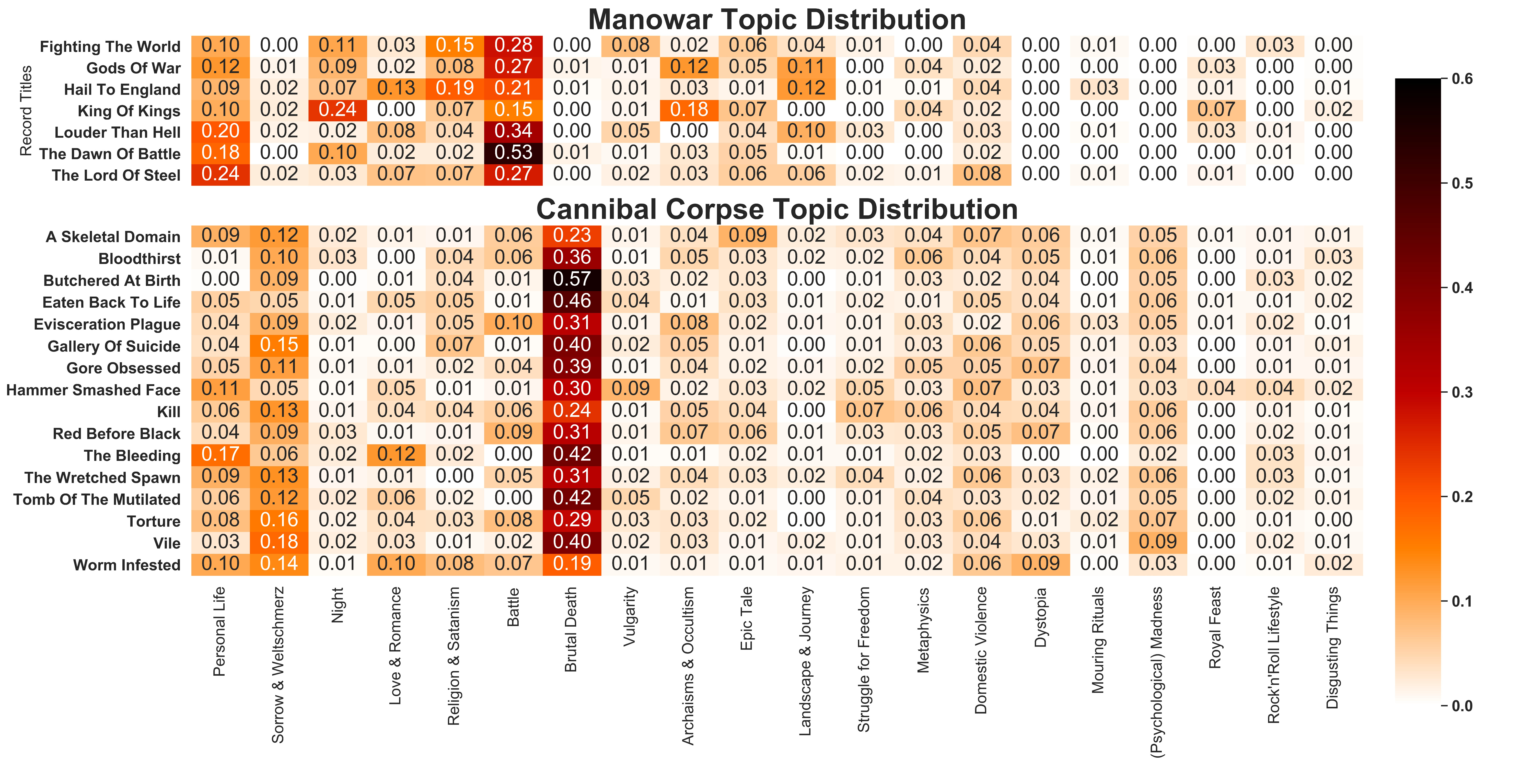}}
  \caption{Comparison of the topic distributions for all included albums by the bands Manowar and Cannibal Corpse showing a prevalence of the topics ‘\emph{battle}’ and ‘\emph{brutal death}’ respectively}
  \label{fig:heatmaps}
\end{figure}

\begin{figure}
  \centering
  \fbox{\includegraphics[width=320pt]{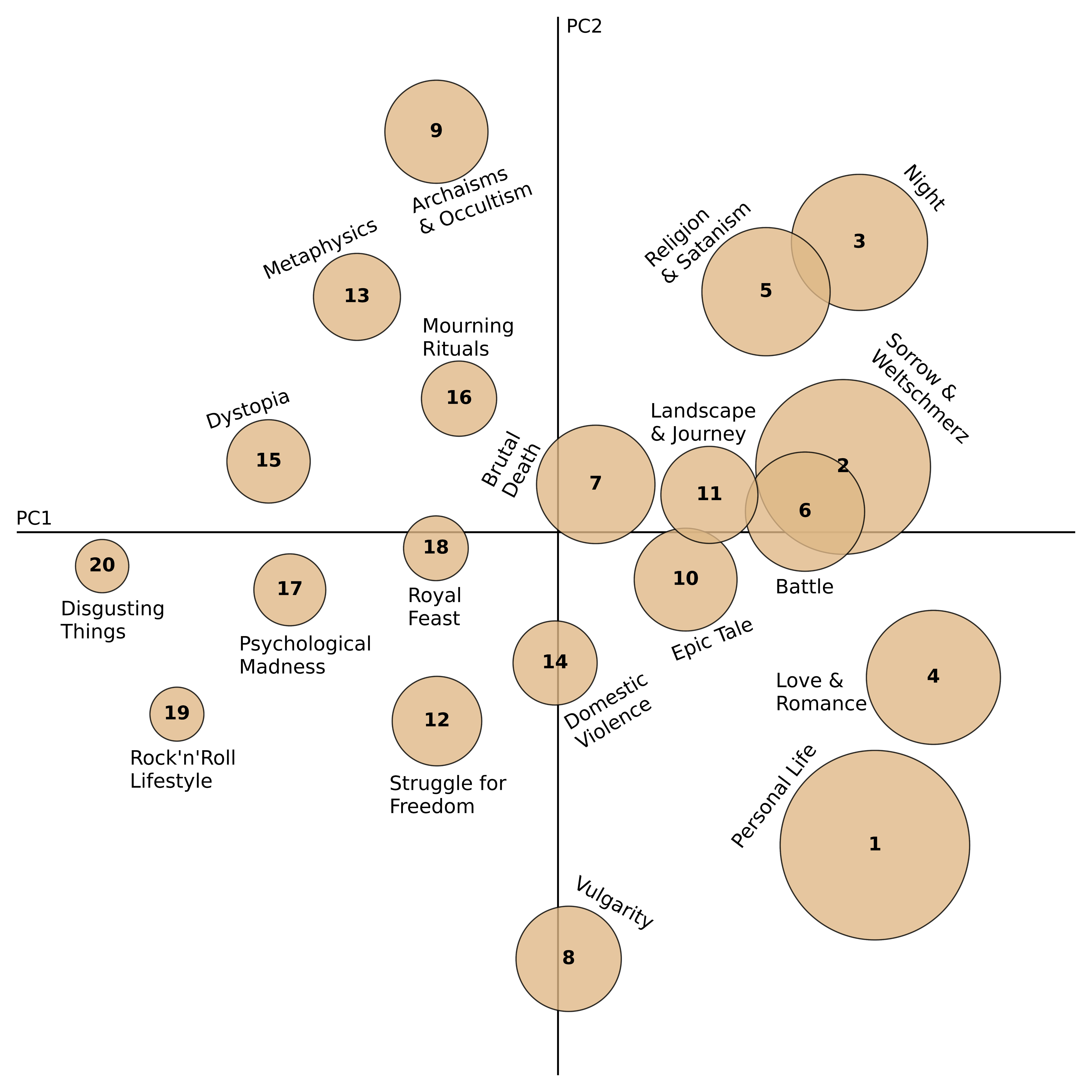}}
  \caption{Topic configuration obtained via multidimensional scaling. The radius of the circles is proportional to the percentage of tokens covered by the topics (topic weight).}
  \label{fig:mds}
\end{figure}

\begin{figure}
  \centering
  \fbox{\includegraphics[width=460pt]{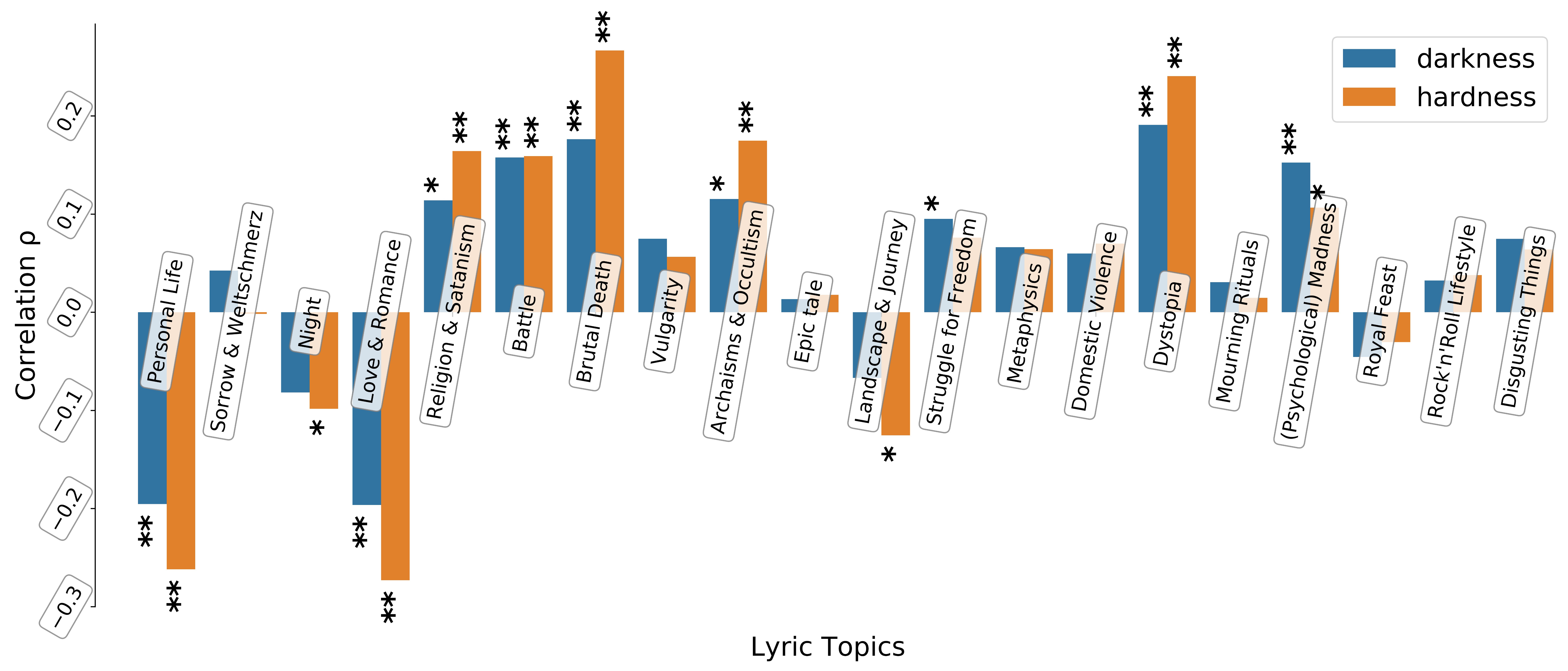}}
  \caption{Correlations between lyrical topics and the musical dimensions \emph{hardness} and \emph{darkness};
 $*$:$~p~<~0.05$, $**$:$~p~<~0.00125$ (Bonferroni-corrected significance level)}
  \label{fig:correlations}
\end{figure}

\begin{multicols}{2}

This is highlighted in detail in Figure \ref{fig:heatmaps}. Here, the topic distributions for two exemplary bands contained within the sample are presented. For these heat maps, data has been aggregated over individual songs showing the topic distribution at the level of albums over a band’s history. The examples chosen illustrate the dependence between textual topics and musical subgenres. For the band Manowar, which is associated with the genre of heavy metal, power metal or true metal, a prevalence of topic \#6 (‘\emph{battle}’) can be observed, while a distinctive prevalence of topic \#7 (‘\emph{brutal death}’) becomes apparent for Cannibal Corpse – a band belonging to the subgenre of death metal.

Within the topic configuration obtained via multidimensional scaling (see Figure \ref{fig:mds}), two latent dimensions can be identified. The first dimension (PC1) distinguishes topics with more common wordings on the right hand side from topics with less common wording on the left hand side. This also correlates with the weight of the topics within the corpus. The second dimension (PC2) is characterized by an contrast between transcendent and sinister topics dealing with \emph{occultism}, \emph{metaphysics}, \emph{satanism}, \emph{darkness}, and \emph{mourning} (\#9, \#3, .\#5, \#13, and \#16) at the top and comparatively shallow content dealing with \emph{personal life} and \emph{Rock’n’Roll lifestyle} using a rather mundane or \emph{vulgar} vocabulary (\#1, \#8, and \#19) at the bottom. This contrast can be interpreted as ‘\emph{otherworldliness / individual-transcending narratives}’ vs. ‘\emph{worldliness / personal life}’.

\subsection{Correlations with Musical Dimensions}
\label{sec:results_corr}

In the final step of our analysis, we calculated the association between the twenty topics discussed above and the two high-level audio features \emph{hardness} and \emph{darkness} using Spearman’s $\rho$. The results are visualized in Figure \ref{fig:correlations} and the $\rho$~values listed in table \ref{tab:topic_table}.

Significant positive associations can be observed between musical \emph{hardness} and the topics ‘\emph{brutal death}’, ‘\emph{dystopia}’, ‘\emph{archaisms \& occultism}’, ‘\emph{religion \& satanism}’, and ‘\emph{battle}’, while it is negatively linked to relatively mundane topics concerning ‘\emph{personal life}’ and ‘\emph{love \& romance}’. The situation is similar for \emph{dark/gloomy} sounding music, which in turn is specifically related to themes such as ‘\emph{dystopia}’ and ‘\emph{(psychological) madness}’. Overall, the strength of the associations is moderate at best, with a tendency towards higher associations for \emph{hardness} than \emph{darkness}. The strongest association exists between \emph{hardness} and the topic ‘\emph{brutal death}’ ($\rho = 0.267$, $p < 0.01$\footnote{The p values fall below the Bonferroni-corrected significance level.}).

\section{Conclusion and Outlook}

Applying the example of metal music, our work examined the textual topics found in song lyrics and investigated the association between these topics and high-level music features. By using LDA and MDS in order to explore prevalent topics and the topic space, typical text topics identified in qualitative analyses could be confirmed and objectified based on a large text corpus. These include e.g. satanism, dystopia or disgusting objects.
It was shown that musical \emph{hardness} is particularly associated with harsh topics like ‘\emph{brutal death}’ and ‘\emph{dystopia}’, while it is negatively linked to relatively mundane topics concerning personal life and love. We expect that even stronger correlations could be found for metal-specific topics when including more genres covering a wider range of \emph{hardness/darkness} values\footnote{The previously established ground truth for the creation of the audio feature models included far more genres. Therefore, it can be assumed that our metal sample covers only a small fraction of the \emph{darkness/hardness} range.}.

Therefore, we suggest transferring the method to a sample including multiple genres. Moreover, an integration with metadata such as genre information would allow for the testing of associations between topics, genres and high-level audio features. This could help to better understand the role of different domains in an overall perception of genre-defining attributes such as \emph{hardness}.

\bibliographystyle{unsrt}  


\end{multicols}
\end{document}